\documentclass[]{interact}

\usepackage{graphicx}
\usepackage{epstopdf}
\usepackage{subfigure}

\usepackage[numbers,sort&compress,merge]{natbib}
\bibpunct[, ]{[}{]}{,}{n}{,}{,}
\renewcommand\bibfont{\fontsize{10}{12}\selectfont}

\theoremstyle{plain}

\theoremstyle{definition}

\theoremstyle{remark}

\begin{document}

\title{Modeling Magnetic Anisotropy of Single Chain Magnets in $\mathbf{|d/J|\geq 1}$ Regime}

\author{
\name{Sumit Haldar\textsuperscript{a}, Rajamani Raghunathan\textsuperscript{a}, Jean-Pascal Sutter\textsuperscript{b} and S. Ramasesha\textsuperscript{a,*}\thanks{*Corresponding author e-mail: ramasesh@sscu.iisc.ernet.in}}
\affil{\textsuperscript{a} Solid State and Structural Chemistry Unit,\\~~Indian Institute of Science, Bangalore - 560012, INDIA.\\\textsuperscript{b}CNRS, LCC, 205 Route de Narbonne, F-31077 Toulouse, France.\\~~Universit\'e de Toulouse, UPS, INPT, LCC, F-31007 Toulouse, France.}
}

\maketitle

\begin{abstract}
Single molecule magnets (SMMs) with single-ion anisotropies $\mathbf d$, comparable to exchange interactions $J$, between spins have recently been synthesized. In this paper, we provide theoretical insights into the magnetism of such systems. We study spin chains with site spins, $s=1$, $3/2$ and $2$ and on-site anisotropy $\mathbf d$ comparable to the exchange constants between the spins. We find that large $\mathbf d$ leads to crossing of the states with different $M_S$ values in the same spin manifold of the $\mathbf d=0$ limit. For very large $\mathbf d$'s we also find that the $M_S$ states of the higher energy spin states descend below the $M_S$ states of the ground state spin manifold. Total spin in this limit is no longer conserved and describing the molecular anisotropy by the constants $D_M$ and $E_M$ is not possible. However, the total spin of the low-lying large $M_S$ states is very nearly an integer and using this spin value it is possible to construct an effective spin Hamiltonian and compute the molecular magnetic anisotropy constants $D_M$ and $E_M$. We report effect of finite sizes, rotations of site anisotropies and chain dimerization on the effective anisotropy of the spin chains.
\end{abstract}

\begin{keywords}
Single Chain Magnets; Anisotropy; Model Hamiltonian
\end{keywords}

\date{\today}

\baselineskip=22pt
\section{\label{sec:introduction}Introduction}
In the area of magnetism, single molecule magnets (SMMs) have attracted wide attention for their promise as qubits in quantum computers \cite{Ladd2010,Leuenberger2001}. Among different materials that are being explored for representing a qubit, SMMs are fascinating because (1) it is easy to chemically control and tailor their molecular structure and properties and (2) atomic arrangement in each qubit is identical. SMMs are metallo-organic complexes containing  transition metal ions (eg.: $Mn^{3+}, Mn^{4+}, Ni^{2+}, Fe^{3+}$ etc.) \cite{Wernsdorfer2002, Andres2002, MarilenaFerbinteanu2006} or lanthanide ions (eg.: $Dy^{3+}, Tb^{3+}$ etc.) \cite{Rinehart2011,Woodruff2013a,Tang2006,Sessoli2009,Ganzhorn2013,Mannini2014} or a combination of both as active spin centers. The exchange pathways between the magnetic centers are provided by simple ligand groups such as $R=(OH)^-,(CN)^-$ and $O^{2-}$. SMMs are characterized by high ground state spin ($S_{g}$) and large uniaxial magnetic anisotropy mainly coming from the spin-orbit interactions within the ion centers. Contribution of magnetic anisotropy to the spin Hamiltonian, for weak anisotropic interaction is given by 
\begin{eqnarray}
\label{eqn:MolAnisH}
\hat{\mathcal{H}}_{aniso}=D_M\left( \hat{S}_z^2 - \frac{1}{3}S(S+1) \right)+E_M\left(\hat{S}_x^2-\hat{S}_y^2 \right)
\end{eqnarray}
\noindent where, $D_M$ is the axial anisotropy and $E_M$ is the transverse anisotropy, S is the total spin of the state and $\hat{S}_x$, $\hat{S}_y$ and $\hat{S}_z$ are the total spin operators. An essential requirement for a high spin molecule to be a SMM is that $D_M < 0$ and $|D_M| >> E_M$. In the absence of any anisotropy, the ground state of the molecule is $(2S_{g}+1)$ fold degenerate corresponding to different orientations of the spin with $M_S=-S_{g}, -S_{g}+1,\dots ,S_{g}$. Negative $D_M$ ensures a magnetic ground state with positive and negative $M_S$ values confined to two different quantum wells. If the energy barrier $E_B$ given by $|D_M| M_S^2$ is sufficiently large, then the magnetized state is trapped in one of the wells, below a blocking temperature, $T_B$. The transverse anisotropy $E_M$ is responsible for tunneling of magnetization from one well to another. The spins can also relax by climbing over the barrier through spin-phonon interactions or the Orbach process at high enough temperatures \cite{Tyryshkin}. Below $T_B$, SMMs exhibit magnetic hysteresis like in bulk ferromagnets and are hence considered as quantum analogues of classical magnets. These properties of SMMs have raised the hope of realizing molecular materials for quantum computing \cite{Leuenberger2001a}, spintronic \cite{Lehmann2009,Bogani2008,Camarero2009} and high density data storage applications. Besides, SMMs are also being pursued to understand the rich physics behind quantum phenomena like quantum tunneling of magnetization (QTM) and quantum coherence \cite{Hill2003, Wernsdorfer1999}. The grand challenge in this field is to raise the temperature below which these molecules can be used as qubits. This requires raising the energy barrier for the spin to crossover from one well to another, which can be achieved by simultaneously large $S_{g}$ and a large negative $D_M$. Theoretical modeling of anisotropy of SMMs is thus necessary to design systems with large anisotropy.

One approach to enhancing $T_B$ is to synthesize single chain magnets (SCMs) in which the metal ions and bridging ligands are arranged on 1-dimensional lattice \cite{RodolpheClerac2002,Tian-FuLiu2003,HitoshiMiyasaka2003,Miyasaka2009,Coulon2006,Coulon2004}. These SCMs can be synthesized with varying chain lengths. The one dimensional nature can help control the arrangement of metal ions, enabling realization of SCMs with large $D_M$ values. During the last decade, a cobalt based SCM \cite{Caneschi2001,Gambardella2002} showing magnetic hysteresis below 4 K analogous to SMMs was synthesized and this led to a flurry of activity for synthesizing SCMs. 
However, the spin glass behavior of SCMs complicates our understanding of the relaxation process in magnetic chains. It has been shown that the anisotropy barrier is known to depend not only on $D_M$, but also on the strength of magnetic exchange $J$ between the spin centers \cite{Feng2012}. Thus, SCMs are considered to be the most suitable choice to realize systems with high effective energy barrier, $U_{eff}$, leading to synthesis of several Co based systems with barriers as high as ($396$K). But, in most of these cases the $D_M$ values are positive rendering the ground state to be non-magnetic; the slow relaxation in these systems is governed by small planar anisotropy $E_M$ in the $X-Y$ plane.

 Mononcuclear SMMs have only one spin center in the molecule and hence the anisotropy can be tuned easily by controlling coordination around the metal ion. In transition metal complexes with many spin centers, presence of large number of bridging ligands makes it harder to control SMM properties. To realize SMMs and SCMs with large anisotropy barriers, two main approaches are being adapted: (1) increasing the number of ion centers and (2) using spin centers with large single-ion anisotropy $\mathbf d$. The former approach has tremendous potential to yield SMMs with large $S_{g}$ and $|D_M|$ simultaneously. However,  the highly symmetric nature of larger high-nuclearity transition metal clusters usually lead to small $D_M$ values, for example in case of $Mn_{19}$, $S_{g}=\frac{83}{2}$ and $D_M$ is very low \cite{Ako2006}. The latter approach of using ions with large $\mathbf d$ is recently being investigated. This has motivated designing SMMs with rare-earth ions which have both large spins and large magnetic anisotropy due to the unquenched orbital angular momentum. This has already resulted in surpassing the $U_{eff}$  values of some of the most well known transition metal based SMMs like $Mn_{12}Ac$. For e.g., the $U_{eff}$ values of $Dy_4$ and $Dy_5$ SMMs are $692 K$ and $528 K$ respectively \cite{Blagg2013}. However, there are key challenges that still remain to be addressed: (1) increasing the strength of superexchange of $4f$ ions, (2) overcoming difficulties in tuning exchange due to high ligand coordination. The relaxation processes in lanthanides with large $U_{eff}$ are considered to be more complex than the thermally activated behavior found in transition metal based SMMs. Apart from the usual QTM and Orbach thermal relaxation processes, a third thermally assisted  quantum resonant tunneling \cite{Friedman1996} in which the ground state relaxes through an excited state has been proposed to understand slow relaxation in these systems. 

Theoretical modeling and prediction of $D_M$ and $E_M$ has been very challenging. Density functional theoretical (DFT) calculations have been used by Pedersen et. al \cite{Postnikov2006} as well as by Neese et. al \cite{Neese1998} to compute anisotropies of SMMs. They obtain the gradient of the potential $\phi(\vec{r})$ and using $\vec{s}\cdot [\vec{p}\times \vec{\nabla}\phi(\vec{r})]$ as a perturbation, compute the anisotropy constants correct to second order. However,  it should be noted that DFT calculations do not conserve total spin of the state or the site-spin. Thus, the ground and the excited states obtained from a DFT calculation do not have spin purity and are contaminated due to the admixture of other spin states. Besides, usually multinuclear SMMs have exchange interaction which have frustrations leading to some intermediate spin ground states \cite{Raghunathan2006}. DFT can not target these states as they are not fully spin polarized. Hence, the $D_M$ and $E_M$ values can not be obtained for a chosen spin state of the system in contrast to experiments that show $D_M$ and $E_M$ are strongly dependent on the total spin of the state. Thus the DFT approaches have been successful in predicting the anisotropies of single-ion magnets rather than the molecular anisotropy of SMMs with multiple spin centers.

In our previous work \cite{Raghunathan2008}, we developed a theoretical approach to compute $D_M$ and $E_M$ for SMMs in any spin eigenstate of the exchange Hamiltonian, treating the anisotropy Hamiltonian as a perturbation. This method is very generic and uses the spin-spin correlations between various sites in a chosen eigenstate of the exchange Hamiltonian. Further, the method relies only upon the single-ion anisotropies $\mathbf d$ as the input parameter and can compute the molecular anisotropy as a function of orientation of the magnetic axes of the individual ions. This method was previously employed to calculate the $D_M$ and $E_M$ parameters of two well known SMMs namely $Mn_{12}Ac$ and $Fe_8$. In these two well known systems the magnetic anisotropy is very weak compared to the strength of magnetic exchange interaction. However, there exist many SMMs \cite{Walsh2014,Bernot2008,Miller2000} in which the single-ion anisotropies are comparable in magnitude to the Heisenberg exchange constant $J$. Since our earlier method is based on a perturbative approach, it can not be applied to systems where $\mathbf d\geq J_{ij}$. This demands that the anisotropy term should be included in the exchange Hamiltonian whose eigenstates are used to compute the molecular anisotropy parameters. The approach that is presented here is very generic and can be applied to compute the anisotropy parameters of any SMMs given the anisotropy tensor of the magnetic centers and spins, irrespective of the relative strength of anisotropy and exchange parameters. In the following section, we present our methodology. In section \ref{sec:RandD}, we discuss the effect of $\mathbf d$ on mixing of spin states. Then we apply this methodology to model spin chains and discuss the effect of $\mathbf d$, scaling of the molecular anisotropy with number of spin sites $N$, orientation of single-ion anisotropy and dimerization on the molecular magnetic anisotropy. We summarize our results in the last section.

\section{\label{sec:methodology}Modeling and Methodology}

The full magnetic Hamiltonian of molecular magnets can be described as a sum of the exchange and anisotropic parts,
\begin{eqnarray}
\label{eqn:HeisenbergAnis}
\hat{\mathcal{H}}^{full}=\hat{\mathcal{H}}_{ex} + \hat{\mathcal{H}}_{aniso}
\end{eqnarray}
The exchange part is given by  
\begin{eqnarray}
\label{eqn:Heisenberg}
\hat{\mathcal{H}}_{ex}=\sum_{<ij>} J_{ij} \hat{s}_i\cdot \hat{s}_j
\end{eqnarray}
where, $J_{ij}$ s are the exchange constants which can be be either ferromagnetic (negative) or antiferromagnetic (positive) between site spins $s_i$ and $s_j$, connected by an exchange pathway involving the ligands. The spins $s$ are all taken to be isotropic and their value is determined by the metal ion. In this study we consider linear spin chains with site spins $s=1$, $3/2$ and $2$.
The general anisotropic Hamiltonian is given by 
\begin{eqnarray}
\hat{\mathcal{H}}_{aniso}^{full}&=&\sum_{i \geq j}\sum_{\alpha, \beta}  \hat{s}_i^\alpha  \lambda_{ij}^{\alpha\beta}  \hat{s}_{j}^{\beta}
\end{eqnarray}
where  $\lambda_{ij}^{\alpha\beta}$ is the anisotropic interaction matrix element for the interactions between spins $i$ and $j$. The origin of $\lambda_{ij}$ can be either spin-orbit or spin dipolar interactions. In metal ions, the spin-dipolar interactions are much smaller than spin-orbit interactions. The anisotropic interactions between two different centers falls off as $\sim R_{ij}^{-3}$, where  $\sim R_{ij}^{-3}$ is the distance between $i$ and $j$ and can be neglected. Thus, it is sufficient to consider anisotropic interactions on same magnetic center, leading to
\begin{eqnarray}
\label{eqn:anisHam}
\hat{\mathcal{H}}_{aniso}&=&\sum_{i}\sum_{\alpha, \beta}  \hat{s}_i^\alpha  \lambda_{i}^{\alpha\beta}  \hat{s}_{i}^{\beta}
\end{eqnarray}
If the magnetic axis of different spin centers differ, then $\hat{\mathcal{H}}_{aniso}$ can be transformed to the laboratory frame, using direction cosines of the local axis $\alpha$ of the magnetic ion. Thus, eqn. \ref{eqn:anisHam} can be rewritten as 
\begin{eqnarray}
\label{eqn:MasterEq}
\hat{\mathcal{H}}_{aniso}&=&\sum_{i} \sum_{l, m} \sum_{\alpha, \beta} \lambda_i^{\alpha\beta} C_i^{l\alpha} C_i^{m\beta} \hat{s}_i^{\alpha} \hat{s}_i^{\beta}
\end{eqnarray}
where $C_i^{l\alpha}$ is the direction cosine of the $l^{th}$ axis of the laboratory frame with the local axis $\alpha$. In all our studies, we consider only diagonal anisotropy at the sites and rotate all the site anisotropies from the transverse to longitudinal direction of the linear spin chain. Thus the microscopic anisotropic Hamiltonian we consider is 
\begin{eqnarray}
\label{eqn:anisodterm}
\hat{\mathcal{H}}_{aniso}&=&\sum_i d_i \hat{s}_{i,z}^2
\end{eqnarray}
This form of $\hat{\mathcal{H}}_{aniso}$ is used in eqn. (\ref{eqn:HeisenbergAnis}) in our studies.
For a given SMM with many spin centers, the $\hat{\mathcal{H}}_{aniso}$ for a total spin state $|n, S\rangle$, can be written as
\begin{eqnarray}
\label{eqn:leftHam}
\hat{\mathcal{H}}_{aniso}&=&\sum_{l,m}  \hat{S}^{l}  \Lambda^{lm}  \hat{S}^{m}
\end{eqnarray}
\noindent where, $\Lambda^{lm}$ is the anisotropy tensor of the molecule, $l$, $m$ = $X,~Y,~Z$ are the laboratory coordinate axes.  Diagonalizing the $3 \times 3$ matrix $\Lambda^{lm}$ to obtain the eigenvalues $\Lambda^X$, $\Lambda^Y$ and $\Lambda^Z$, eqn. \ref{eqn:anisHam} can be recast as
\begin{eqnarray}
\hat{\mathcal{H}}_{aniso}&=&\Lambda^{X} (\hat{S}^X)^2 + \Lambda^{Y} (\hat{S}^Y)^2 + \Lambda^{Z} (\hat{S}^Z)^2
\end{eqnarray}
where $|\Lambda^Z|$ is assumed to be the largest eigenvalue and the corresponding eigenvector defines the z-axis of the SMM. Since, the quantity of interest is the energy level splitting of the spin state $|n, S\rangle$, we can impose the trace of the matrix $\Lambda^{lm}$ to be zero and define
\begin{eqnarray}
\label{eqn:DEterm}
D_M&=&\Lambda^{Z}-\frac{1}{2}\left( \Lambda^{X}+\Lambda^{Y}\right)\nonumber\\
E_M&=&\frac{1}{2}\left( \Lambda^{X}-\Lambda^{Y}\right)
\end{eqnarray}

and write $\hat{\mathcal{H}}_{aniso}$ in the state $|n, S \rangle$ as
\begin{eqnarray}
\hat{\mathcal{H}}_{aniso}=D_M\left[ (\hat{S}^Z)^2 - \frac{1}{3}S(S+1) \right]+E_M\left[(\hat{S}^X)^2-(\hat{S}^Y)^2 \right]
\end{eqnarray}
The first term in the Hamiltonian in the above equation does not commute with the microscopic $\hat{\mathcal{H}}_{aniso}$ in eqn.(\ref{eqn:anisodterm}). Thus even when the anisotropies of all the site spins are oriented along the z-direction, the molecular anisotropy picks up the transverse or the tunneling anisotropic term $E_M$ in eqn. (\ref{eqn:DEterm}).
In ref. \citep{Raghunathan2008}, it is shown in detail how $D_M$ and $E_M$ can be obtained for an eigenstate $|n, S\rangle$ from the microscopic Hamiltonian in eqn. \ref{eqn:Heisenberg} using first order perturbation theory. When the single ion anisotropies are comparable to or larger than the exchange strength ($|d/J| \geq 1.0$), the above method is not suitable and the full Hamiltonian has to be employed to obtain the energies of the system. In general neither $\hat{S}^2$ nor $\hat{S}_z$ is conserved and one needs to deal with the full Hamiltonian.
In our studies, we have considered two different cases, namely ($i$) $|d/J| < 1.0$ and ($ii$) $|d/J| \geq 1.0$. In ($i$), The eigenstates of the exchange Hamiltonian are only weakly perturbed by the anisotropic interactions. We can use the treatment in \cite{Raghunathan2008} to obtain the molecular anisotropy constants $D_M$ and $E_M$. In ($ii$) since $|\mathbf d|$ is of the same magnitude as $|J_{ij}|$, we diagonalize the full Hamiltonian. The full Hamiltonian no longer conserves total spin and total spin is not a good quantum number. In case the anisotropy of the individual sites are all aligned along the z-axis of the laboratory frame then total ${S_z}$ is conserved. In this case, we can solve the full Hamiltonian in different $M_S$ sectors. However, in general the full Hamiltonian needs to be diagonalized in the full Fock space of the spin Hamiltonian, which for N spins of spin $s$ has the dimension $(2s + 1)^N$.
In principle for strong anisotropy, we can not define the molecular anisotropy constants $D_M$ and $E_M$ as the spin of the states are not defined. However, on computing the expectation value of $\hat{S}_{tot}^2$ in the low-lying states, we find that it is approximately close to $S(S+1)$, where $S$ is an integer. Thus, we can assume this nearest integer to be the spin $S$ of the state and proceed to compute the molecular anisotropy matrix $\Lambda^{lm}$. To obtain the matrix elements of $\Lambda^{lm}$, we first compute the matrix elements $\langle S, M_S| \hat{\mathcal{H}}_{aniso}| S, M_S ^\prime \rangle$, where $\hat{\mathcal{H}}_{aniso}$ is from eqn. (\ref{eqn:leftHam}), using standard spin algebra with the spin of the state taken to be $S$. This matrix elements involve the unknown matrix elements of $\Lambda^{lm}$. The same matrix elements can also be computed by taking $\hat{\mathcal{H}}_{aniso}$ from eqn. (\ref{eqn:MasterEq}). This involves computing the relevant spin-spin correlations $\langle S, M_S| \hat{S}_i^\alpha \hat{S}_i^\beta | S, M_S^\prime \rangle$. Equating the two yields a set of linear algebraic equations in $\Lambda^{lm}$. In general there are $(2S + 1)^2$ equations corresponding to the $(2S + 1)^2$ matrix elements $\hat{\mathcal{H}}_{aniso},  \langle S, M_S| \hat{\mathcal{H}}_{aniso}| S, M_S ^\prime \rangle$. The number of constants $\Lambda^{lm}$ are only 9 as $l,m=X,Y,Z$. If the spin is strictly conserved, we can use any of the 9 equations from  $(2S + 1)^2$ equations as Wigner-Eckart theorem strictly holds and $\Lambda^{lm}$ values obtained do not depend on the choice of equations. Since, $S$ is not strictly defined for the strong anisotropy case, the Wigner-Eckart theorem does not strictly apply and this appears to be a bottleneck. However, we have verified that the Wigner-Eckart theorem is approximately valid even when $S$ is not strictly conserved. This allows us to determine the matrix elements $\Lambda^{lm}$ and the corresponding eigenvalues $\Lambda^X$, $\Lambda^Y$  and $\Lambda^Z$
 from which we can obtain $D_M$ and $E_M$. 

\section{\label{sec:RandD}Results and Discussion}
\subsection{\label{ssec:Mixing}Mixing of Spin States in the Strong Anisotropy Limit}

\begin{figure*}
    \includegraphics[width=4in]{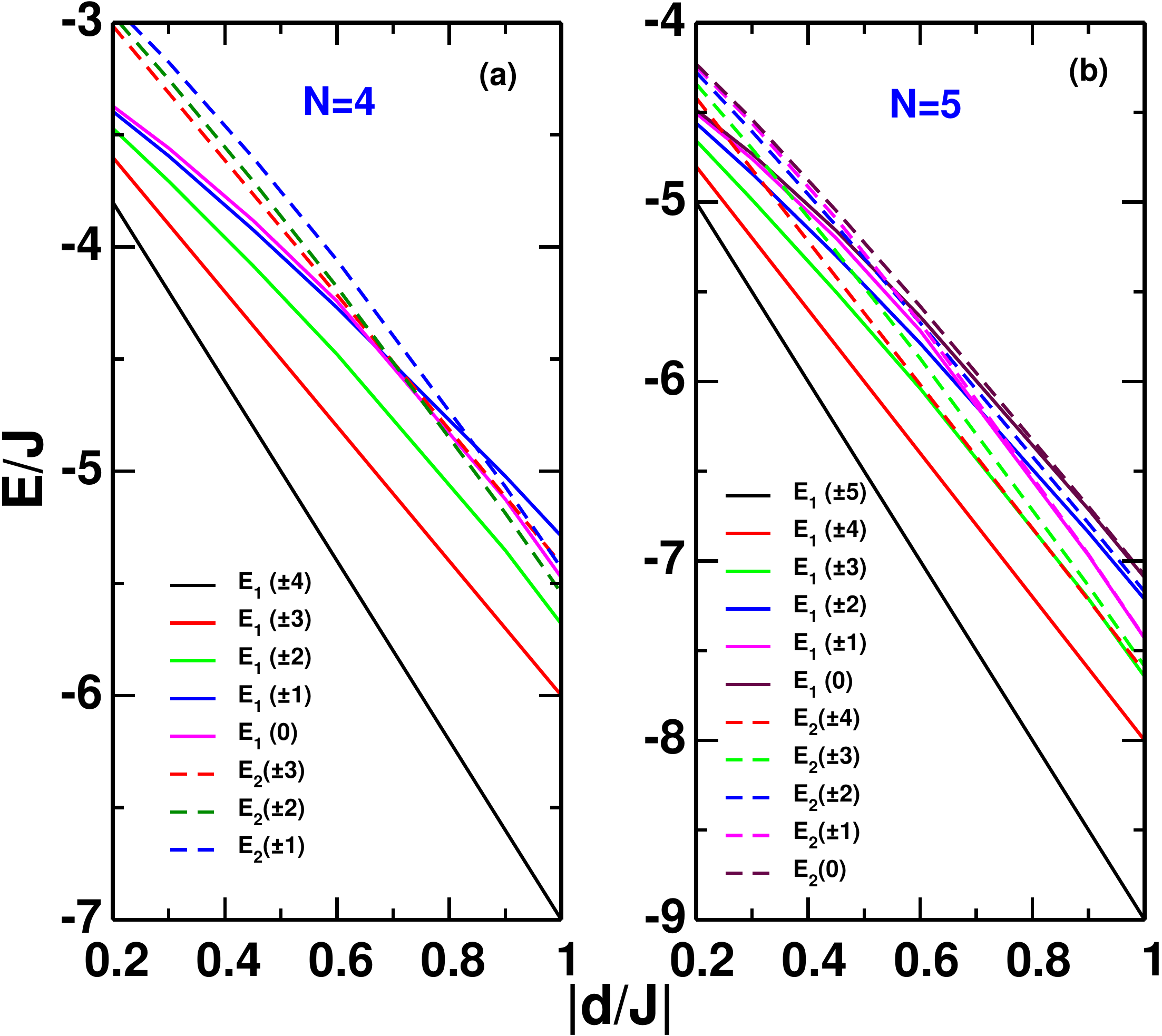}
      \caption{\label{fig:EvsdiN} Energy {\it vs.} $|d/J|$ plot of levels corresponding to the ground state $GS$ ($E_1$) and the first excited state ($E_2$) spin manifolds in case of $s=1$ for (a) $N~=~4$ and (b) $N~=~5$ sites. The corresponding $M_S$ values are given in the brackets. Energy levels of only those excited states that cross the ground state spin manifold are shown.}
\end{figure*}

\begin{figure*}
    \includegraphics[width=5in]{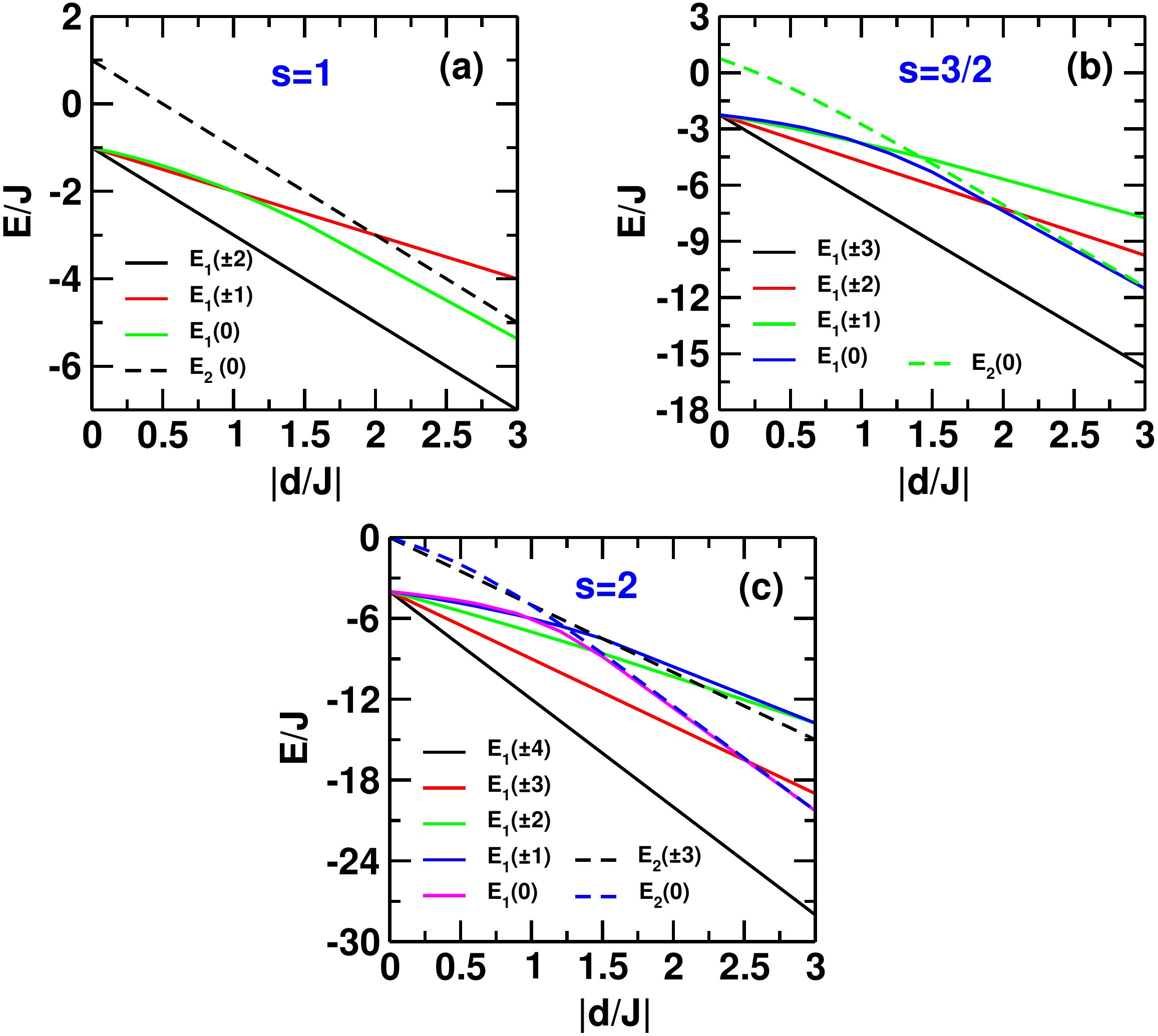}
      \caption{\label{fig:Evsdisi}Energy {\it vs.} $|d/J|$ plot of levels corresponding to the ground state $GS$ ($E_1$) and the first excited state ($E_2$) spin manifolds in case of (a) $s=1$, (b) $s=3/2$ and (c) $s=2$ for a dimer. The corresponding $M_S$ values are given in the brackets. Energy levels of only those excited states that cross the ground state spin manifold are shown.}
\end{figure*}

\begin{table*}
\caption{Energy, E/J and total spin of the state, $S_{tot}$ for different $M_S$ states obtained from the expectation value of the $\langle\hat{S}^2\rangle$ operator, for $|d/J|=0.09$, $0.3$ and $0.9$.}
\label{tab:Ssqexpvalues}
\setlength{\tabcolsep}{2.0pt}
\begin{tabular}{|c|c|c|c|c|c|c|c|c|c|c|} \hline
 & \multicolumn{2}{|c|} {$|d/J|$=0.09} &  &  & 
 \multicolumn{2}{|c|} {$|d/J|$=0.3} &  & &
 \multicolumn{2}{|c|} {$|d/J|$=0.9}   \\  
\cline{2-3} \cline{6-7} \cline{10-11}
$|M_S|$ & E/J & $S_{tot}$ &  & $|M_S|$ & E/J & $S_{tot}$ & & $|M_S|$ & E/J & $S_{tot}$ \\ \hline
5.0 & -4.449 & 5.000 & & 5.0 & -5.499 & 5.000 & & 5.0 & -8.500 & 5.000\\ 
4.0 & -4.359 & 5.000 & & 4.0 & -5.199 & 5.000 & & 4.0 & -7.600 & 5.000\\
3.0 & -4.291 & 4.997 & & 3.0 & -4.986 & 4.967 & & 4.0 & -7.218 & 4.000\\
2.0 & -4.243 & 4.995 & & 2.0 & -4.841 & 4.931 & & 3.0 & -7.210 & 4.365\\
1.0 & -4.214 & 4.993 & & 4.0 & -4.818 & 4.000 & & 3.0 & -7.138 & 3.792\\ 
0.0 & -4.205 & 4.992 & & 1.0 & -4.758 & 4.887 & & 1.0 & -6.969 & 3.038\\
4.0 & -3.978 & 4.000 & & 0.0 & -4.731 & 4.877 & & 1.0 & -6.964 & 2.922\\
3.0 & -3.935 & 3.998 & & 3.0 & -4.702 & 3.980 & & 2.0 & -6.840 & 4.239\\ 
2.0 & -3.905 & 3.995 & & 2.0 & -4.606 & 3.953 & & 2.0 & -6.789 & 3.736\\
1.0 & -3.886 & 3.992 & & 1.0 & -4.560 & 3.889 & & 0.0 & 6.712 & 3.672\\ 
0.0 & -3.880 & 3.991 & & 0.0 &  4.539 & 3.892 & & 0.0 & 6.712 & 3.672\\ \hline
\end{tabular}
\end{table*}

\begin{figure}
    \includegraphics[width=5in]{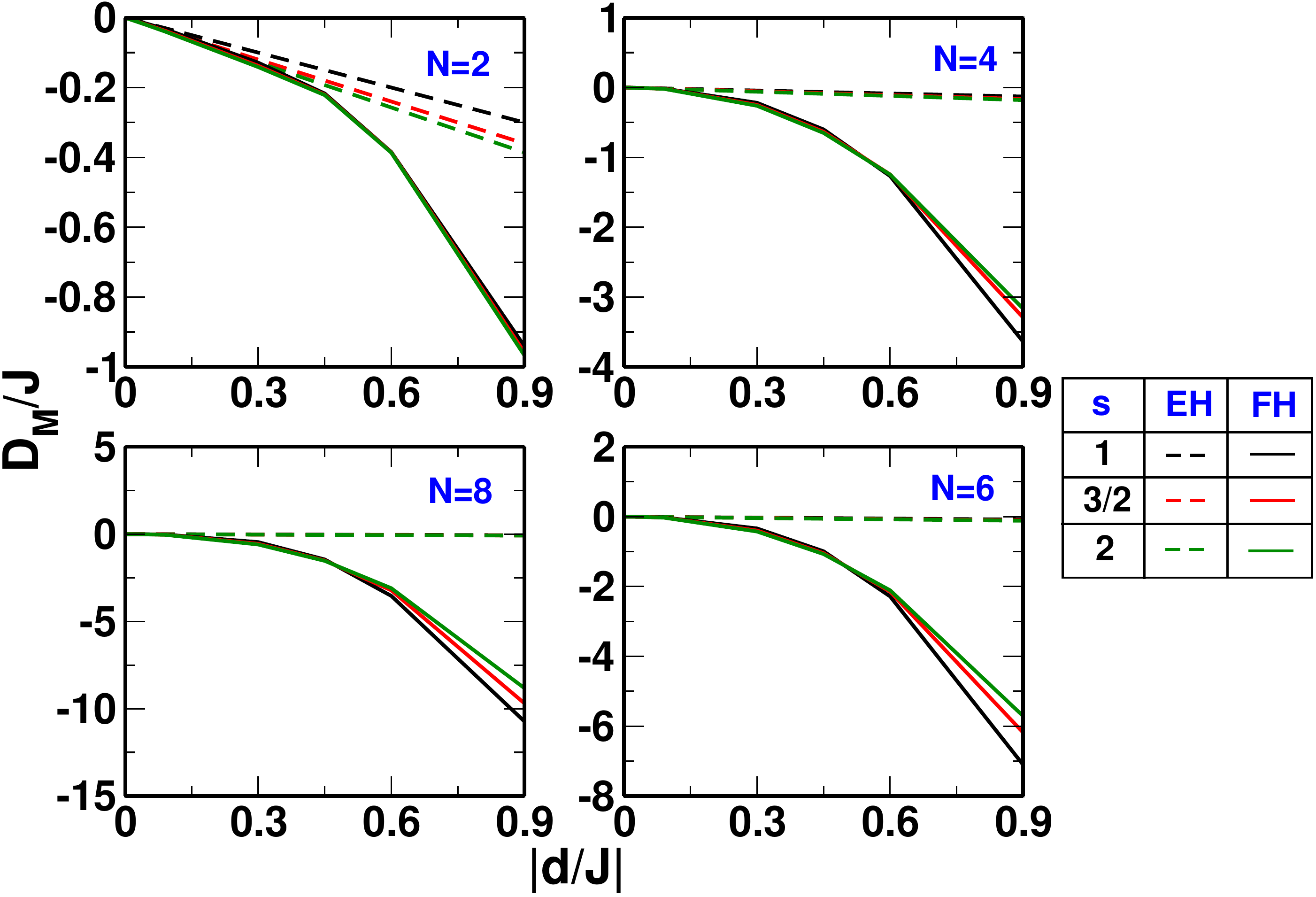}
      \caption{\label{fig:dimerDvsd} Effect of $|d/J|$ on the molecular anisotropy ($D_M$) of  $s=1$, $s=3/2$, $s=2$ spin dimers. Dashed and solid lines represent $D_M$ values computed using spin correlations obtained from exchange Hamiltonian (EH) and full Hamiltonian (FH) respectively.}
\end{figure}

We have studied the evolution of low-lying energy levels as a function of $|d/J|$ in spin chains of different lengths ($N$) and different site spins ($s$), using exact diagonalization of the Hamiltonian in different total $M_S$ sectors. We have assumed that the ferromagnetic exchange is the same for all nearest neighbour spin pairs and that the local anisotropy is aligned along the laboratory $Z$ axis. In this case, the total $M_S$ is a good quantum number. Large on-site anisotropy leads to crossing of the energy levels with different $|M_S|$ values as shown in the figure \ref{fig:EvsdiN}. For small anisotropy, we note that the energies of the states decrease with decreasing $|M_S|$ values. However, for large $|d/J|$, we note that the lowest energy  $M_S=0$ state descends below the lowest energy $|M_S|=\pm1$ state for $N=4$ and as the anisotropy is increased further, the second lowest energy level corresponding to $|M_S|=3$ and $|M_S|=1$ also descend below the $M_S=0$ level. For $N=5$, there are more level crossings. For small values of $|d/J|$, there is a crossing of the lowest states with $|M_S|=1$ and $M_S=0$ followed by crossing of the second lowest state with $|M_S|=4$ and the lowest $|M_S|=1$ and the $M_S=0$ states. 

\noindent These crossings can be understood by considering the $N=2$ case. In this case, the ground state has total spin $S=2$ when $\mathbf d=0$. Upon turning on $\mathbf d$, the states with $|M_S|=2$, $1$ and $0$ split with energies of the state increasing as we decrease $|M_S|$ (see eqn. 1). However, for a larger $\mathbf d$, the state with $M_S=0$ descends below the state with $|M_S|=1$. This is due to the fact that the $M_S=0$ state can be obtained from the $m_{s,1}=1$, $m_{s,2}=-1$ and $m_{s,1}=-1$, $m_{s,2}=1$ combinations of the z-component of the site spins. However, the $M_S=1$ state can be obtained from $m_{s,1}=1$, $m_{s,2}=0$ and $m_{s,1}=0$, $m_{s,2}=1$ states. As we see, the $M_S=0$ state is stabilized more than the $M_S=1$ state for large $\mathbf d$ as the site contribution to the anisotropy energy is larger for the $M_S=0$ state, since both the site spins have a non-zero $z$-component of spin. We also note that the energy level crossings occur at smaller values of $\mathbf d$ and the number of energy level crossings also increase, with increasing chain length. 

\noindent We thus see that for describing the large anisotropy situation, we can not define a molecular anisotropy parameter $D_M$ in the Hamiltonian given by eqn. (\ref{eqn:MolAnisH}). The magnetic properties require a knowledge of the full eigenvalue spectrum of the microscopic Hamiltonian. However, the expectation value of the total spin operator, $\langle \hat{S}^2 \rangle$ in different states (within the same spin manifold when $\mathbf d=0$) deviate only slightly from the integer values (see Table \ref{tab:Ssqexpvalues}), when the anisotropy is turned on.

\noindent We have also studied spin chains with site spin 3/2 and 2 and we note that there are similar level crossings and the number of level crossing also increase with site spin as shown in fig. \ref{fig:Evsdisi} for a 2-site system. The qualitative nature of level crossings does not depend upon the topology of the system as we see similar effect of $\mathbf d$ on energy levels in rings (see fig. S1 in Supporting Information (SI)). For different site spins, the crossing between states within the manifold occurs at nearly the same $|d/J|$ value. However, the level crossing between multiplets occurs for progressively lower values of $|d/J|$ with increasing site spins.

\noindent We calculated the $D_M$ and $E_M$ values for dimers of $s=1$, $3/2$ and $2$ using two different methods. In the first perturbative approach, the correlation functions required to obtain $D_M$ and $E_M$ are computed using the eigenstates of the exchange only Hamiltonian (EH). In the second, the correlation functions are calculated from the full Hamiltonian (FH), which includes exchange and anisotropic interactions. In this case, the spin of the states is assumed to be the nearest integer $S$ obtained from the expectation value of $\langle \hat{S}^2 \rangle$ equated to $S(S+1)$. In figure \ref{fig:dimerDvsd} we have plotted $D_M/J$ against $d/J$ and we see that the $D_M$ computed using the two methods agree reasonably for $d/J\leqslant0.4$. With increasing chain length the agreement between the two methods shifts to lower $d/J$ values, for all site spins. 

\begin{figure*}
    \includegraphics[width=6cm]{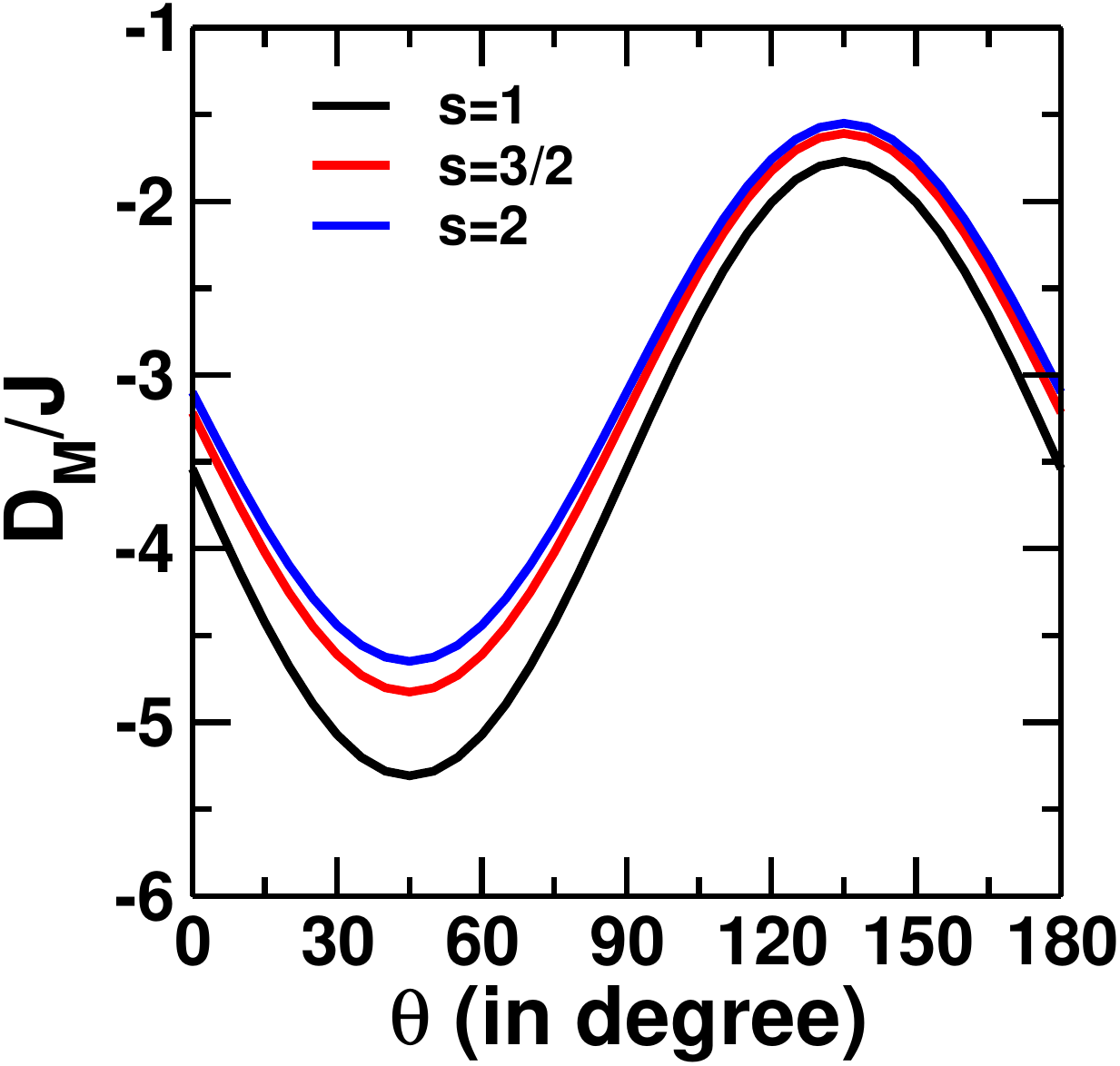} \hspace{0.3in} \includegraphics[width=6.2cm]{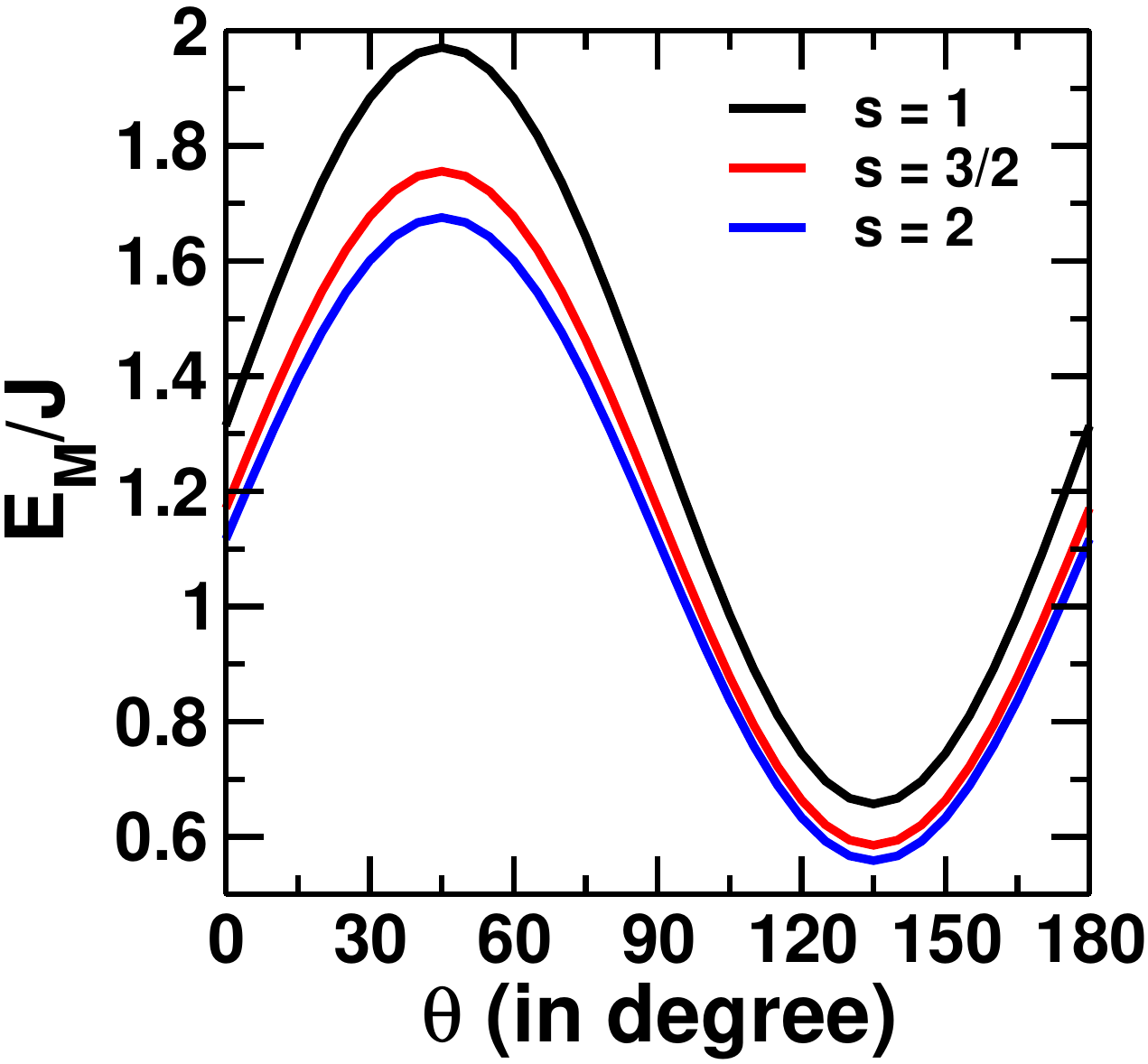} 
      \caption{\label{fig:D vs Euler angle theta for spin}Variation of $D_M$ and $E_M$ as a function of single-ion rotation for an eight site chain ($N=8$) with $s=1$, 3/2 and $2$. The angle between the molecular $Z$ and the single-ion $z$ is given by $\theta$. The other two Euler angles $\phi$ and $\psi$ are fixed at $0^o$. The anisotropy of the ion sites are taken to be $d=-0.6J$.}
\end{figure*}

\subsection{\label{ssec:ion_rot}Rotation of anisotropy axis of site spins}
In all the studies reported hitherto, we assumed that the anisotropic $z$-axis to be perpendicular or transverse to the axis of the chain, taken to be $x$-axis. In this section we explore the effect of rotating the local anisotropy axis around the $y$-axis which tilts the anisotropy towards the chain axis and for $\pi/2$ rotation the anisotropy is along the chain axis. This corresponds to the Euler angles ($\theta, \phi, \psi$) given by ($0, \theta, 0$). We have computed $D_M$ and $E_M$ as a function of the rotation angle $\theta$ using the full Hamiltonian eigenstates for computing the correlation functions. This is shown in figure \ref{fig:D vs Euler angle theta for spin} for the three systems corresponding to $s=1$, 3/2 and $2$ and for a chain length of 8 sites. We find that $D_M$ is negative in all cases and can be empirically fitted to 

\begin{eqnarray}
D_M(\theta)=D_{M_0}(3cos^2(\theta-\delta)-1)+C
\end{eqnarray}

\noindent where, $\delta$ is $45^o$ in all cases, $D_{M_0}$ and $C$ are -1.18 and -2.97 for $s=1$, -1.08 and -2.7 for $s=3/2$ and -1.02 and -2.6 for $s=2$. We see that $|D_M/J|$ is maximum for $\theta=45^o$ and minimum for  $\theta=135^o$. It is not obvious why for $\theta=45^o$ $D_M$ is a maximum as $D_M$ depends both on exchange interactions and site anisotropies. It is also largest for the $s=1$ system. $E_M/J$ also shows a sinusoidal behaviour. For $D_M/J$ maximum $E_M/J$ is also maximum, although $E_M/J$ is always positive and $D_M/J$ is always negative. This implies that the tunneling rate increases for large $D_M/J$. Thus large $D_M/J$ does not imply that the blocking temperature of magnetization is high due to fast magnetic relaxations. We see similar kind of nature for ring also (see fig. S2 in SI).


\subsection{\label{ssec:Chains}Anisotropy of Magnetic Chains}

\begin{figure*}
    \includegraphics[width=\textwidth]{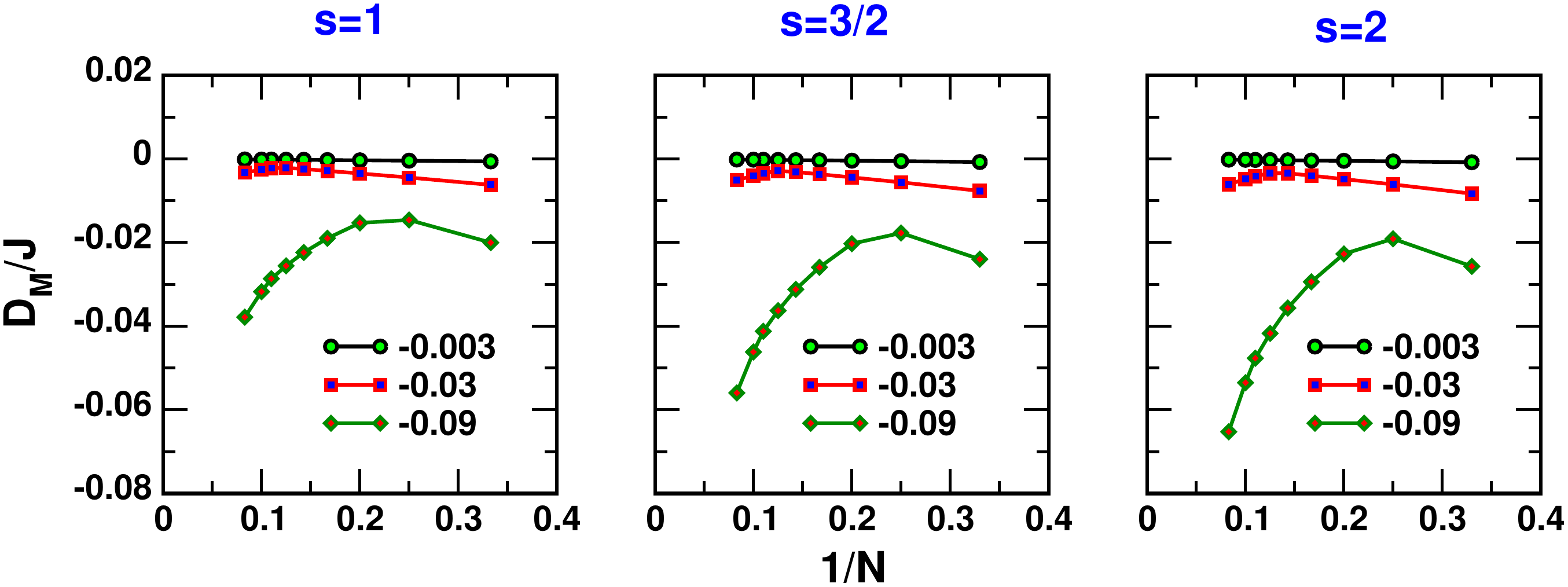}
    \caption{\label{fig:DMvs1byN_WA}Effect of chain length on the molecular anisotropy shown as $D_M$ as a function of inverse system size, $1/N$ for $d=$-0.003, -0.03 and 0.09J.}
\end{figure*}

\begin{figure*}
    \includegraphics[width=\textwidth]{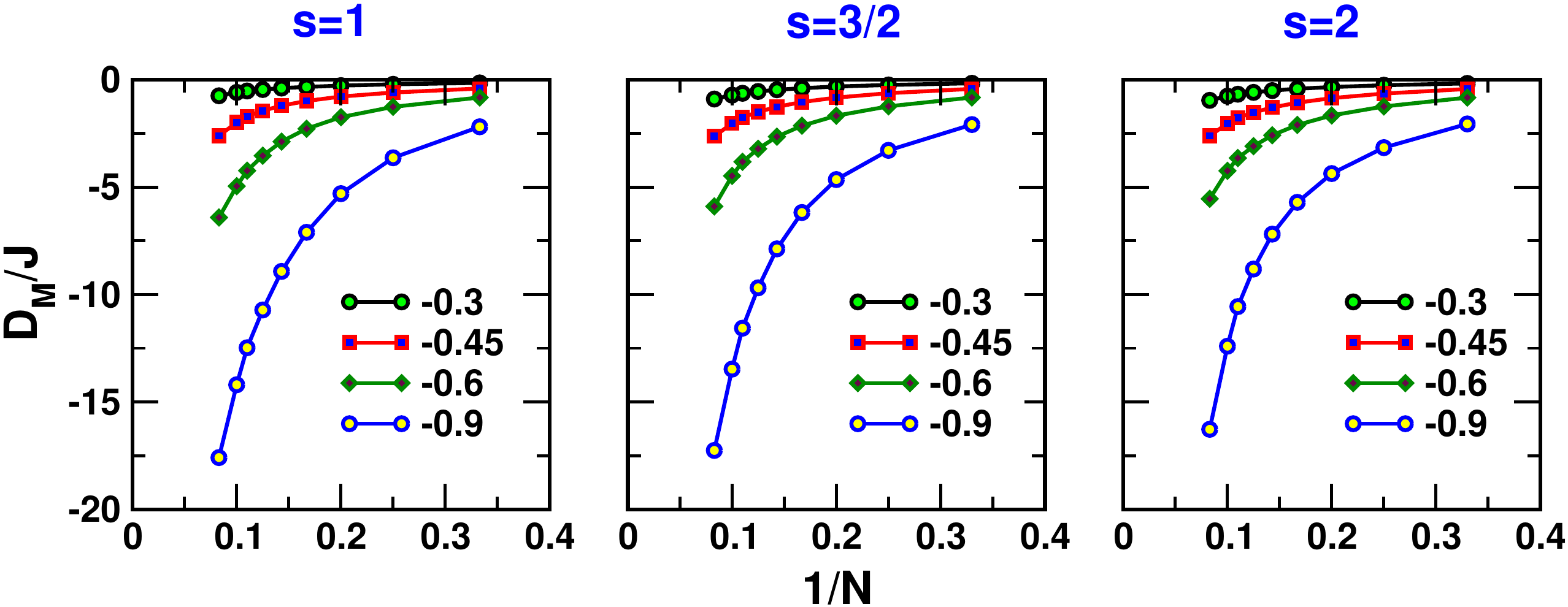}
    \caption{\label{fig:DMvs1byN_SA}Effect of chain length on the molecular anisotropy shown as $D_M$ as a function of inverse system size, $1/N$ for $d=$-0.3, -0.45, -0.6 and -0.9J.}
\end{figure*}

We now discuss the effect of system size on the anisotropy parameters by studying magnetic chains upto 12 sites. The molecular magnetic anisotropy parameters are calculated by using the eigenstates of the full Hamiltonian. Shown in figures \ref{fig:DMvs1byN_WA} and \ref{fig:DMvs1byN_SA} are the dependence of $D_M$ on inverse system size $1/N$ for site spins- 1, 3/2 and 2 and for ring (see figures S3 and S4 in SI) in weak and strong on-site anisotropy parameters respectively. First, we will discuss the weak anisotropy limit where $\mathbf |d/J| ~ \leqslant 0.09$. In this limit, for all the three values of the site spins, we notice that the $|D_M|$ value initially decreases and then increases with increasing $N$. This turn-over occurs for smaller $N$ with increasing $|d/J|$ (fig. \ref{fig:DMvs1byN_WA}). Furthermore, the corresponding $D_M$ values are also higher for larger site spins and larger $|\mathbf d|$ values. These show that larger $D_M$ can be realized by large site anisotropy, $|\mathbf d|$ or large site spin $s$. However, $D_M$ shows a decreasing trend with increasing chain length, $N$ for extremely small site anisotropy $d$. In the strong exchange limit or $0.3 \leqslant \mathbf |d/J| \leqslant 0.9$, our computed $|D_M|$ values increase exponentially with the system size, $N$.

\noindent It is interesting to note that, for a chain length $N$, our computed $D_M$ values in some cases are far larger than the maximum value from simple tensor sum method. For example, with $\mathbf d~=~-0.9~J$ and $N~=~10$ sites, simple tensor summation would predict $D_M$ of $-9~J$, however, our computed $D_M$ values are much larger than this value. In case of SCMs it has been shown previously \cite{Miyasaka2009} that the effective potential barrier for spin flip is the sum of molecular anisotropy barrier created by the magnetic anisotropy of the ions $\Delta_A$ and the barrier created by the spin correlations due to magnetic exchange interactions ($\Delta_C$) for a spin to flip from positive to negative $m_s$. So, it costs an energy of $J$ for the terminal spins to flip, as it is bonded only to one spin center. But, the interior spins are bonded to neighbours, which costs an energy $2J$ for these spins to flip, thus increasing the barrier for magnetization relaxation. The $D_M$ parameter that we obtain from our calculation is the effective anisotropy that also includes the contributions due to the exchange interaction $J$ of the correlated spin-chain. Hence, the computed total $D_M$ values of a chain of length $N$ surpass even the sum of the single-ion contributions. 



\subsection{\label{ssec:dimerization}Effect of Dimerization on the Magnetic Anisotropy}

\begin{figure}
    \includegraphics[width=6cm]{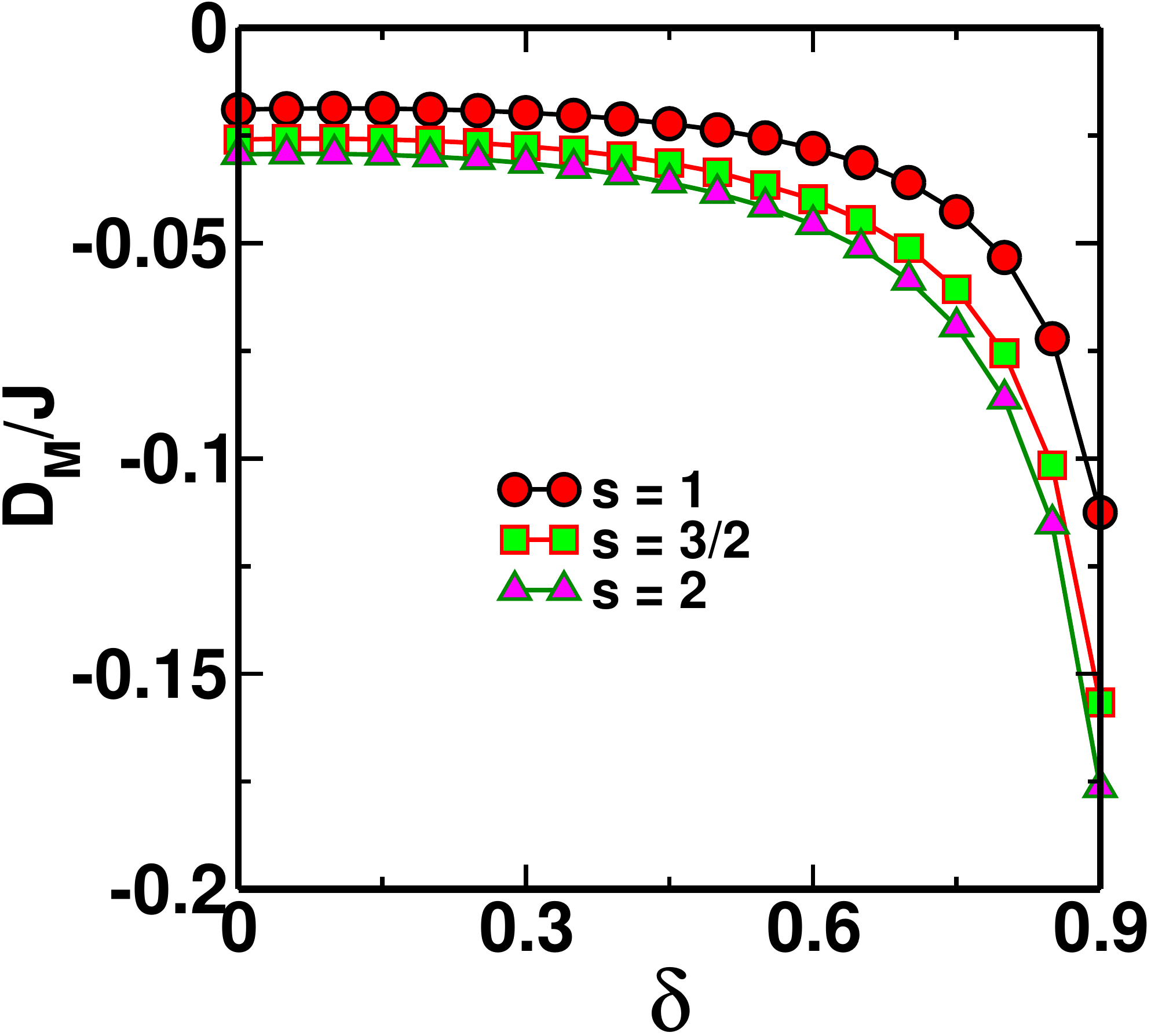}\\\vspace{0.3in}
    \includegraphics[width=5.75cm]{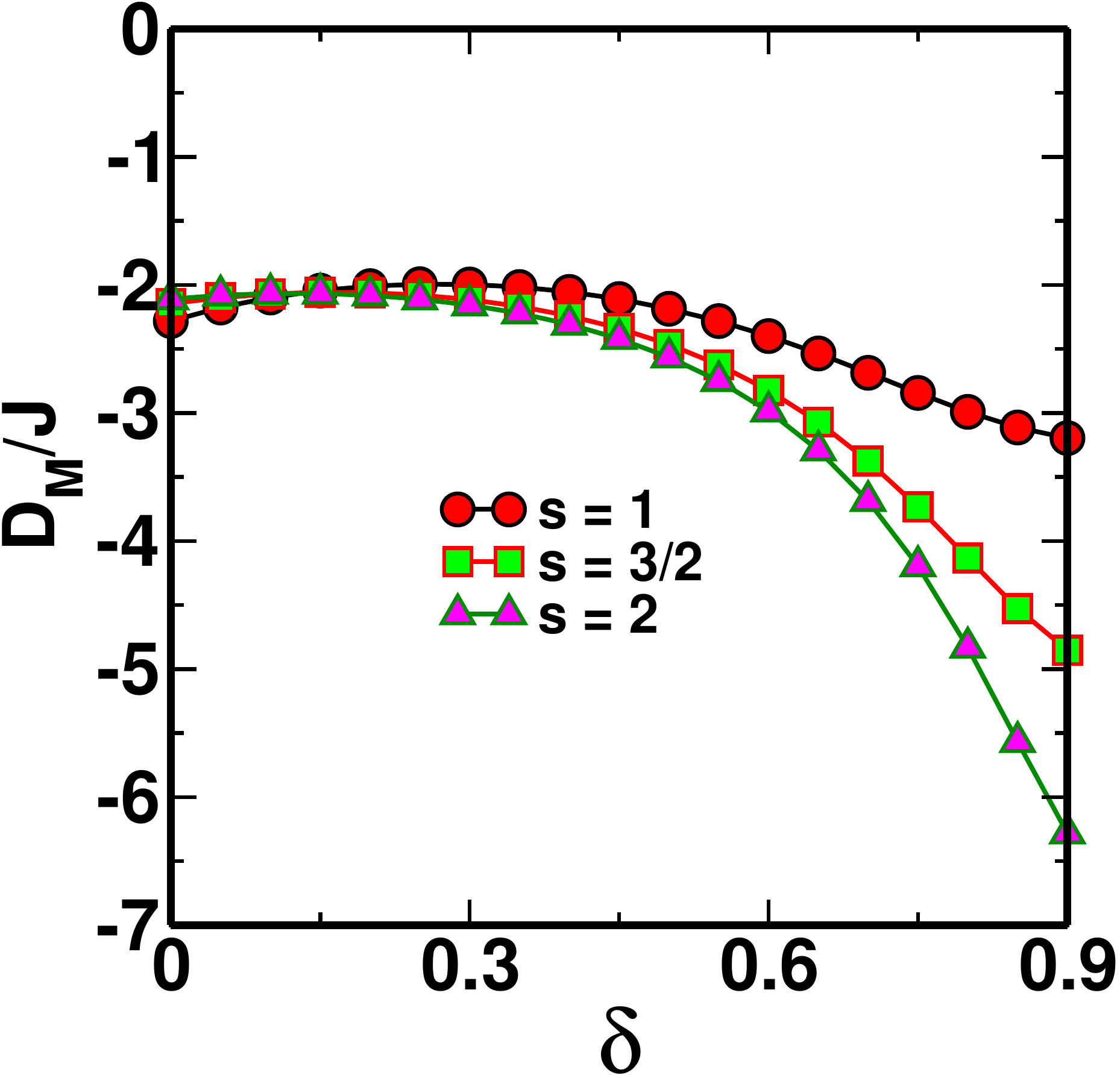}   
    \caption{\label{fig:dimerization}Effect of dimerization on the molecular anisotropy parameter $D_M$ on 6 site spin chains of $s=1$, $3/2$ and $2$ for $|d/J|=0.09$ (top) and $0.6$ (bottom).}
\end{figure}

Spin chains are generally susceptible to spin-Peierls distortion leading to dimerization of the chain in which the exchange constant between successive pairs of spin alternate between $(1+\delta)J_o$ and $(1-\delta)J_o$, where $\delta$ is the extent of dimerization. In fig. \ref{fig:dimerization} we show the variation of $D_M$ as a function of $\delta$, for chains of 6-sites with site spin $s=1$, $3/2$ and $2$ respectively. When $\delta=0.0$, the chain is undimerized and for $\delta=1.0$ the chain is made up of three non-interacting dimers. In the dimer limit the eigenstates are triply degenerate and the degenerate nature of the eigenstates makes the computation of $D_M$ values challenging and requires resolving the degeneracy by forming linear combinations of the degenerate eigenstates. In our study, we limit our discussion to $\delta$ values between 0 and 0.9 for the sake of computational simplicity. We find that for both $|d/J|=0.09$ and $|d/J|=0.6$, the molecular anisotropy parameter increases with $\delta$, but even at $\delta=0.9$ we do not regain the dimer picture. We also notice that, even in presence of dimerization, the $D_M$ values are higher for higher site spins, consistent with our results discussed in the previous section.

\section{\label{sec:conclusion}Conclusions}

We have studied the magnetic anisotropy of a spin chain, given the on-site anisotropy, both in the weak ($|d/J| << 1$) and strong ($|d/J|\approx1$) anisotropy limits. In the strong anisotropy limit, the anisotropic exchange interaction should be included in the exchange Hamiltonian to obtain the eigenstates in which spin-spin correlation functions are calculated for computing the molecular $D_M$ and $E_M$ parameters. We note that strong $|d/J|$ leads to breaking the spin symmetries and even for only diagonal site anisotropies, the lowest energy levels with $M_S=0$ descend below some lowest $M_S\neq0$ energy levels. However, large $M_S\neq0$ levels still have lowest energies ordered with increasing $|M_S|$ values and we may define molecular anisotropy parameters for these states. Besides, the spin expectation values of these states are nearly integers and correspond to the ferromagnetic ground state. We note that the computation of $D_M$ and $E_M$ using correlation functions only from the eigenstates of the exchange Hamiltonian deviates very strongly from those computed from the eigenstates of the full Hamiltonian. Hence, for all $|d/J|$ values we have computed the molecular anisotropy parameters using the full Hamiltonian. Our studies reveal that $|D_M/J|$ and $|E_M/J|$ show a sharp increase with increasing $|d/J|$ and rotating the axis of anisotropy from transverse to longitudinal direction gives a sinusoidal variation in $|D_M/J|$ and $|E_M/J|$. For large negative $D_M/J$ we also have large $E_M/J$ which leads us to believe that magnetization relaxation will be rapid. Increase in chain length of the magnetic chain also leads to sharp increase in magnetic anisotropy, so does dimerization of spin chains.

\section{\label{sec:ack}Acknowledgements}

\noindent SR and JPS thank the Indo-French Centre (IFCPAR/CEFIPRA) for the support. SR thanks DST for support through various grants and INSA for a fellowship. 

\bibliographystyle{tfo}
\bibliography{Sumit_Anisotropy_paper}

\begin{thebibliography}{37}
\providecommand{\url}[1]{\texttt{#1}}
\providecommand{\urlprefix}{URL }

\bibitem{Ladd2010}
T.D. Ladd, F. Jelezko, R. Laflamme, Y. Nakamura, C. Monroe and J.L. O'Brien,
  Nature  \textbf{464} (7285), 45--53 (2010).

\bibitem{Leuenberger2001}
M.N. Leuenberger and D. Loss,  Nature  \textbf{410} (6830), 789--793 (2001).

\bibitem{Wernsdorfer2002}
W. Wernsdorfer, N. Aliaga-Alcalde, D.N. Hendrickson and G. Christou,  Nature
  \textbf{416} (March), 406--409 (2002).

\bibitem{Andres2002}
H. Andres, R. Basler, A.J. Blake, C. Cadiou, G. Chaboussant, C.M. Grant, H.U.
  G{\"{u}}del, M. Murrie, S. Parsons, C. Paulsen, F. Semadini, V. Villar, W.
  Wernsdorfer and R.E.P. Winpenny,  Chemistry - A European Journal  \textbf{8}
  (21), 4867--4876 (2002).

\bibitem{MarilenaFerbinteanu2006}
M. Ferbinteanu, T. Kajiwara, K.Y. Choi, H. Nojiri, A. Nakamoto, N. Kojima, F.
  Cimpoesu, Y. Fujimura, S. Takaishi and M. Yamashita,  Journal of the American
  Chemical Society  \textbf{128} (28), 9008--9009 (2006).

\bibitem{Rinehart2011}
J. Rinehart and J. Long,  Chem. Sci.  \textbf{2} (11), 2078--2085 (2011).

\bibitem{Woodruff2013a}
D.N. Woodruff, R.E.P. Winpenny and R.A. Layfield \textbf{113} (7), 5110--5148
  (2013).

\bibitem{Tang2006}
J. Tang, I. Hewitt, N.T. Madhu, G. Chastanet, W. Wernsdorfer, C.E. Anson, C.
  Benelli, R. Sessoli and A.K. Powell,  Angewandte Chemie - International
  Edition  \textbf{45} (11), 1729--1733 (2006).

\bibitem{Sessoli2009}
R. Sessoli and A.K. Powell \textbf{253} (19-20), 2328--2341 (2009).

\bibitem{Ganzhorn2013}
M. Ganzhorn, S. Klyatskaya, M. Ruben and W. Wernsdorfer,  Nature Nanotechnology
   \textbf{8} (3), 165--169 (2013).

\bibitem{Mannini2014}
M. Mannini, F. Bertani, C. Tudisco, L. Malavolti, L. Poggini, K. Misztal, D.
  Menozzi, A. Motta, E. Otero, P. Ohresser, P. Sainctavit, G.G. Condorelli, E.
  Dalcanale and R. Sessoli,  Nature Communications  \textbf{5}, 4582 (2014).

\bibitem{Tyryshkin}
A.M. Tyryshkin, S.A. Lyon, A.V. Astashkin and A.M. Raitsimring,  Physical
  Review B  pp. 12--15 (2003).

\bibitem{Leuenberger2001a}
M.N. Leuenberger and D. Loss,  Nature  \textbf{410} (6830), 789--793 (2001).

\bibitem{Lehmann2009}
J. Lehmann, A. Gaita-Ari{\~{n}}o, E. Coronado and D. Loss,  Journal of
  Materials Chemistry  \textbf{19} (12), 1672--1677 (2009).

\bibitem{Bogani2008}
L. Bogani and W. Wernsdorfer,  Nature materials  \textbf{7} (3), 179--186
  (2008).

\bibitem{Camarero2009}
J. Camarero and E. Coronado,  Journal of Materials Chemistry  \textbf{19} (12),
  1678 (2009).

\bibitem{Hill2003}
S. Hill, R. Edwards, N. Aliaga-Alcalde and G. Christou,  Science  \textbf{302}
  (5647), 2--6 (2003).

\bibitem{Wernsdorfer1999}
W. Wernsdorfer,  Science  \textbf{284} (5411), 133--135 (1999).

\bibitem{RodolpheClerac2002}
R. Cl{\'{e}}rac, H. Miyasaka, M. Yamashita and C. Coulon,  Journal of the
  American Chemical Society  \textbf{124} (43), 12837--12844 (2002).

\bibitem{Tian-FuLiu2003}
T.F. Liu, D. Fu, S. Gao, Y.Z. Zhang, H.L. Sun, G. Su and Y.J. Liu,  Journal of
  the American Chemical Society  \textbf{125} (46), 13976--13977 (2003).

\bibitem{HitoshiMiyasaka2003}
H. Miyasaka, R. Cl{\'{e}}rac, K. Mizushima, K.I. Sugiura, M. Yamashita, W.
  Wernsdorfer and C. Coulon,  Inorganic Chemistry  \textbf{42} (25), 8203--8213
  (2003).

\bibitem{Miyasaka2009}
H. Miyasaka, M. Julve, M. Yamashita and R. Cl{\'{e}}rac,  Inorganic Chemistry
  \textbf{48} (8), 3420--3437 (2009).

\bibitem{Coulon2006}
C. Coulon, H. Miyasaka and R. Cl{\'{e}}rac,  Structure and Bonding
  \textbf{122}, 163--206 (2006).

\bibitem{Coulon2004}
C. Coulon, R. Cl{\'{e}}rac, L. Lecren, W. Wernsdorfer and H. Miyasaka,
  Physical Review B - Condensed Matter and Materials Physics  \textbf{69} (13),
  132408 (2004).

\bibitem{Caneschi2001}
A. Caneschi, D. Gatteschi, N. Lalioti, C. Sangregorio, R. Sessoli, G. Venturi,
  A. Vindigni, A. Rettori, M.G. Pini and M.A. Novak,  Angewandte Chemie -
  International Edition  \textbf{40} (9), 1760--1763 (2001).

\bibitem{Gambardella2002}
P. Gambardella, A. Dallmeyer, K. Maiti, M.C. Malagoli, W. Eberhardt, K. Kern
  and C. Carbone,  Nature  \textbf{416} (6878), 301--304 (2002).

\bibitem{Feng2012}
X. Feng, J. Liu, T.D. Harris, S. Hill and J.R. Long,  Journal of the American
  Chemical Society  \textbf{134} (17), 7521--7529 (2012).

\bibitem{Ako2006}
A.M. Ako, I.J. Hewitt, V. Mereacre, R. Cl{\'{e}}rac, W. Wernsdorfer, C.E. Anson
  and A.K. Powell,  Angewandte Chemie - International Edition  \textbf{45}
  (30), 4926--4929 (2006).

\bibitem{Blagg2013}
R.J. Blagg, L. Ungur, F. Tuna, J. Speak, P. Comar, D. Collison, W. Wernsdorfer,
  E.J.L. McInnes, L.F. Chibotaru and R.E.P. Winpenny,  Nature chemistry
  \textbf{5} (8), 673--678 (2013).

\bibitem{Friedman1996}
J.R. Friedman, M.P. Sarachik, R. Ziolo and J. Tejada,  Physical Review Letters
  \textbf{76} (20), 3830--3833 (1996).

\bibitem{Postnikov2006}
A.V. Postnikov, J. Kortus and M.R. Pederson,  Physica Status Solidi (B) Basic
  Research  \textbf{243} (11), 2533--2572 (2006).

\bibitem{Neese1998}
F. Neese and E.I. Solomon,  Inorganic chemistry  \textbf{37} (26), 6568--6582
  (1998).

\bibitem{Raghunathan2006}
R. Raghunathan, J.P. Sutter, L. Ducasse, C. Desplanches and S. Ramasesha,
  Physical Review B  \textbf{73} (10), 104438 (2006).

\bibitem{Raghunathan2008}
R. Raghunathan, S. Ramasesha and D. Sen,  Physical Review B  \textbf{78} (10),
  104408 (2008).

\bibitem{Walsh2014}
J.P.S. Walsh, S. Sproules, N.F. Chilton, A.L. Barra, G.A. Timco, D. Collison,
  E.J.L. McInnes and R.E.P. Winpenny,  Inorganic Chemistry  \textbf{53} (16),
  8464--8472 (2014).

\bibitem{Bernot2008}
K. Bernot, J. Luzon, R. Sessoli, A. Vindigni, J. Thion, S. Richeter, D.
  Leclercq, J. Larionova and A. {Van Der Lee},  Journal of the American
  Chemical Society  \textbf{130} (5), 1619--1627 (2008).

\bibitem{Miller2000}
J.S. Miller,  Inorganic Chemistry  \textbf{39} (20), 4392--4408 (2000).

\end{thebibliography}

\end{document}


\preprint{APS/Anisotropy}

\title{Modeling Magnetic Anisotropy of Single Chain Magnets in $\mathbf{|d/J|\geq 1}$ Regime}

\author{%
Sumit Haldar\affmark[1], Rajamani Raghunathan\affmark[1], Jean-Pascal Sutter\affmark[2], S. Ramasesha\affmark[1]\\
\affaddr{\affmark[1]Solid State and Structural Chemistry Unit,\\
 Indian Institute of Science, Bangalore - 560012, INDIA. }\\
\affaddr{\affmark[2]CNRS, LCC, 205 Route de Narbonne, F-31077 Toulouse, France.\\
Universit\'e de Toulouse, UPS, INPT, LCC, F-31007 Toulouse, France.}\\
  \vspace{3.5in} {\centerline {\bf \huge  Supporting Information}}
}



 \maketitle

\renewcommand\thefigure{S\arabic{figure}}

\begin{figure*}
    \includegraphics[width=6in]{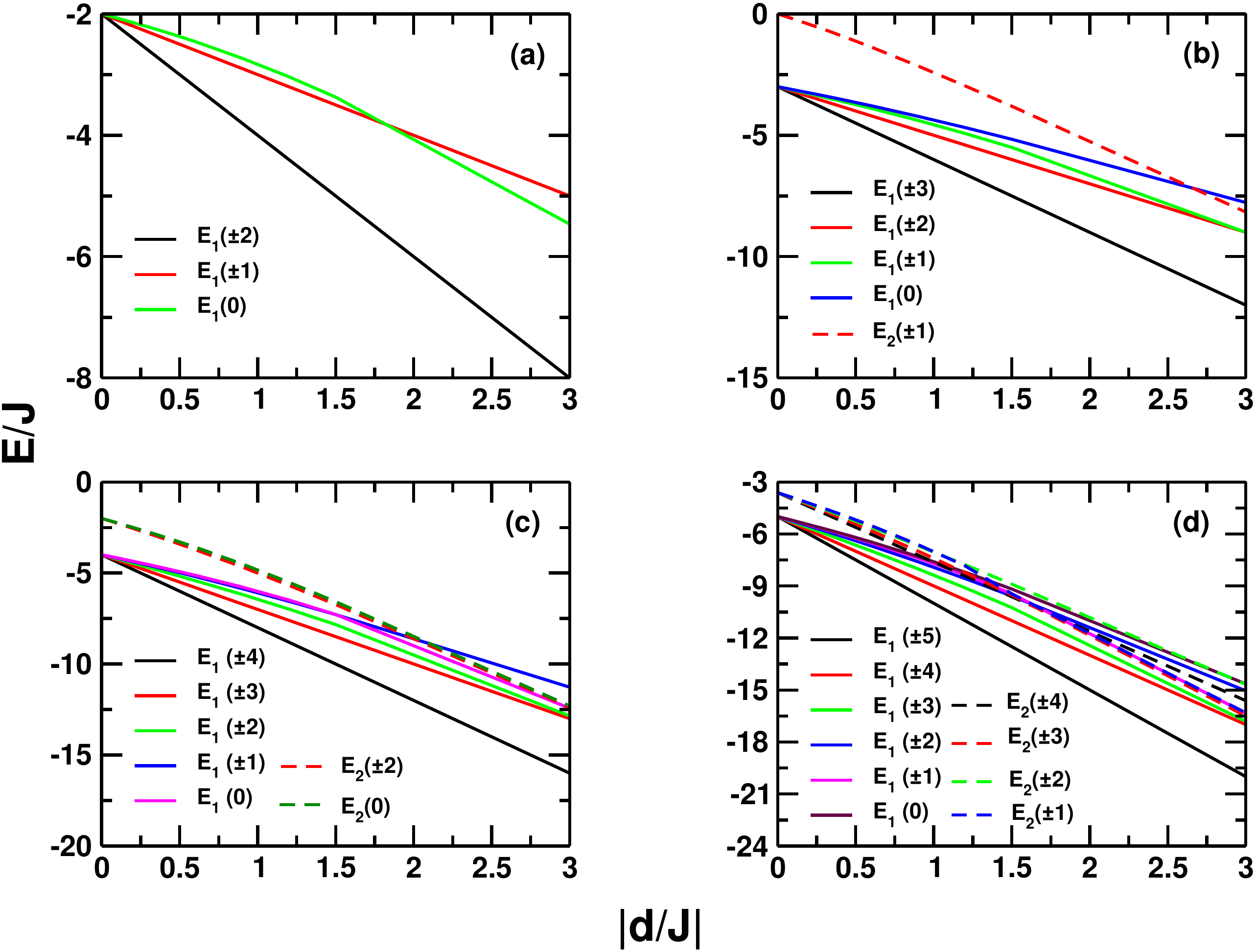}
      \caption{\label{fig:EvsdiN} Energy {\it vs.} $|d/J|$ plot of levels corresponding to the ground state $GS$ ($E_1$) and the first excited state ($E_2$) spin manifolds in case of $s=1$ for (a) $N~=~2$, (b) $N~=~3$, (c) $N~=~4$ and (d)$N~=~5$ site ring. The corresponding $M_S$ values are given in the brackets. Energy levels of only those excited states that cross the ground state spin manifold are shown.}
\end{figure*}



\begin{figure*}
    \includegraphics[width=6cm]{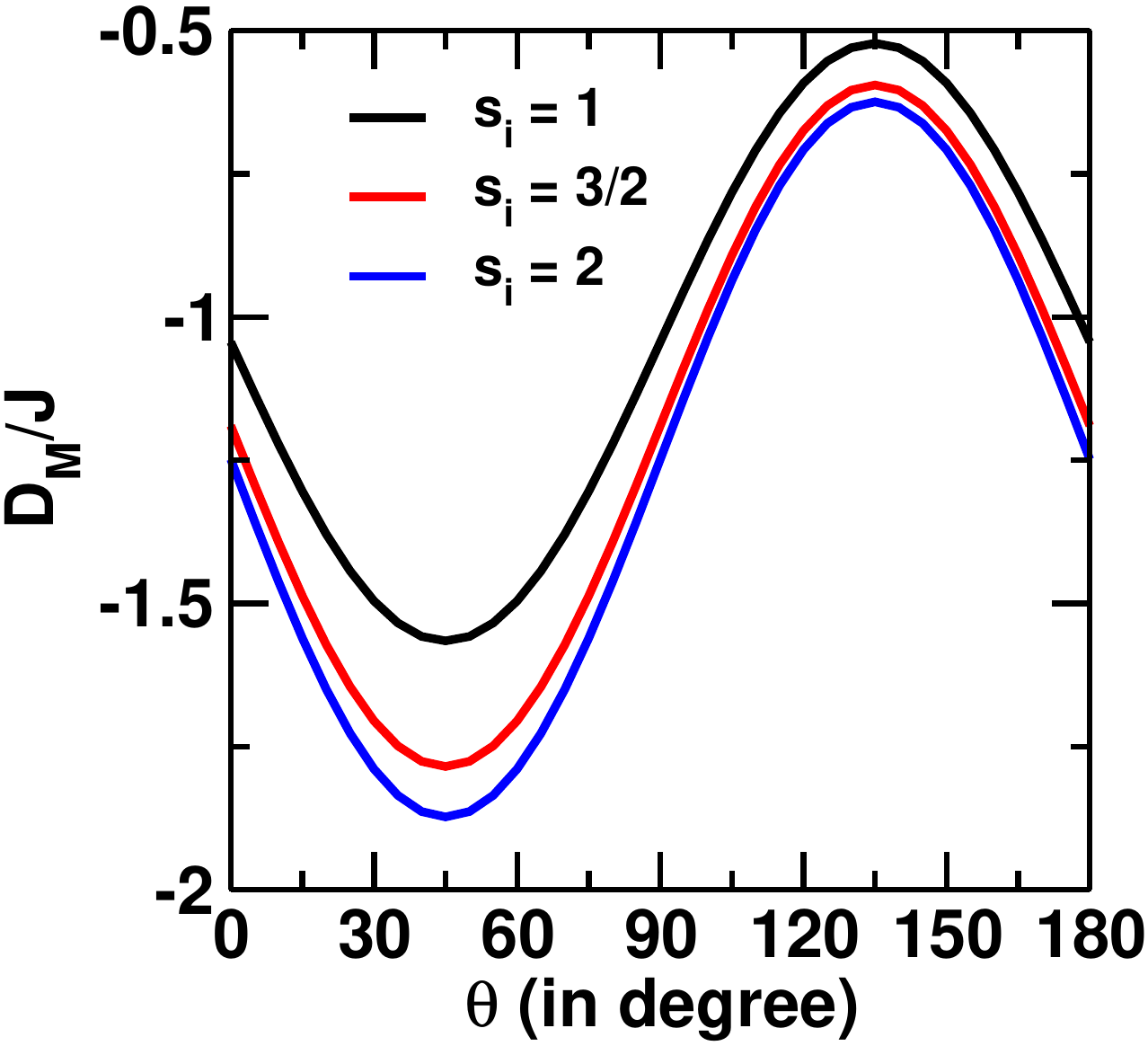} \hspace{0.3in} \includegraphics[width=5.9cm]{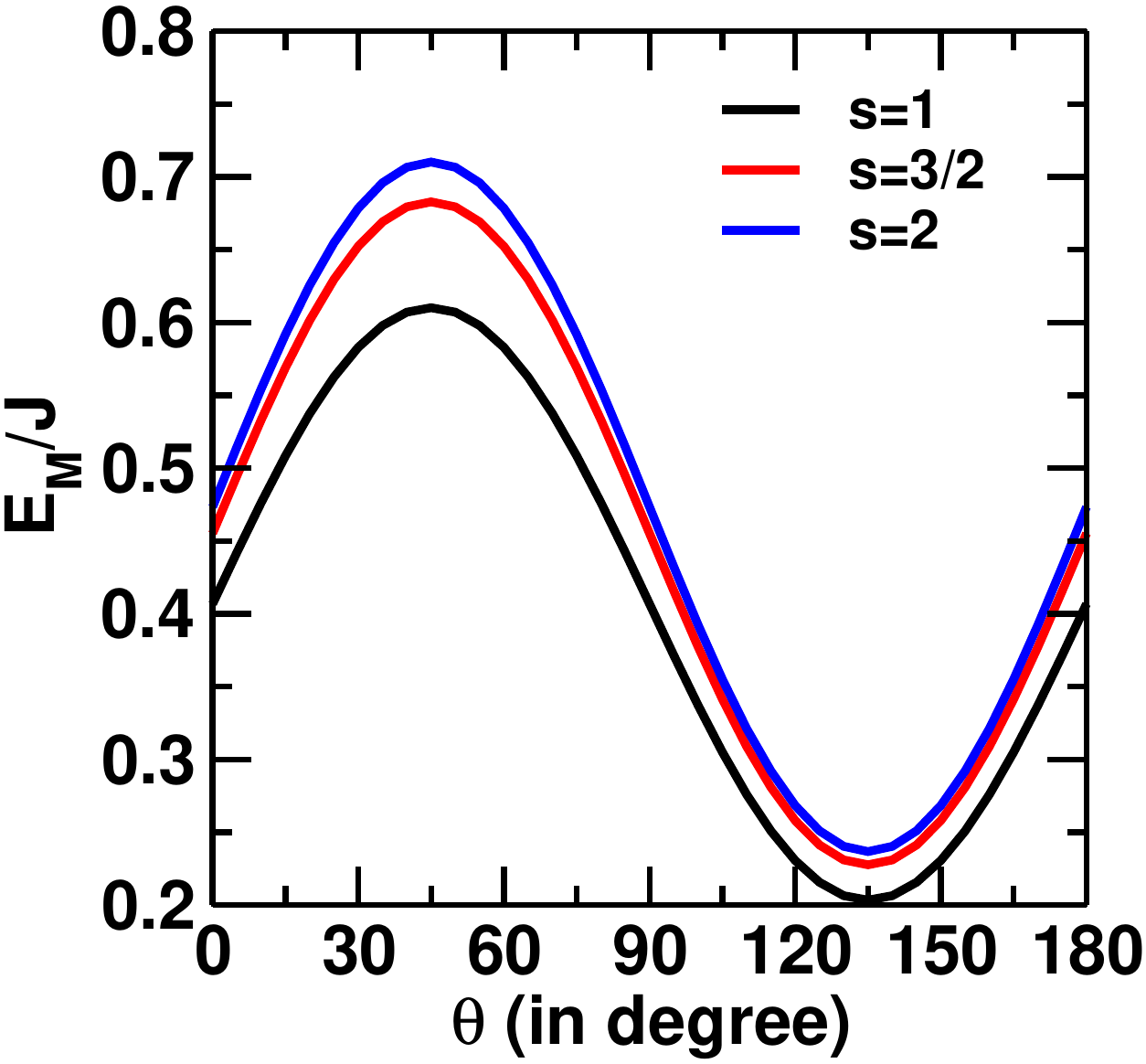} 
      \caption{\label{fig:D vs Euler angle theta for spin}Variation of $D_M$ and $E_M$ as a function of single-ion rotation for an eight site ring ($N=8$) with $s=1$, 3/2 and $2$. The angle between the molecular $Z$ and the single-ion $z$ is given by $\theta$. The other two Euler angles $\phi$ and $\psi$ are fixed at $0^o$. The anisotropy of the ion sites are taken to be $d=-0.6J$.}
\end{figure*}

\begin{figure*}
    \includegraphics[width=\textwidth]{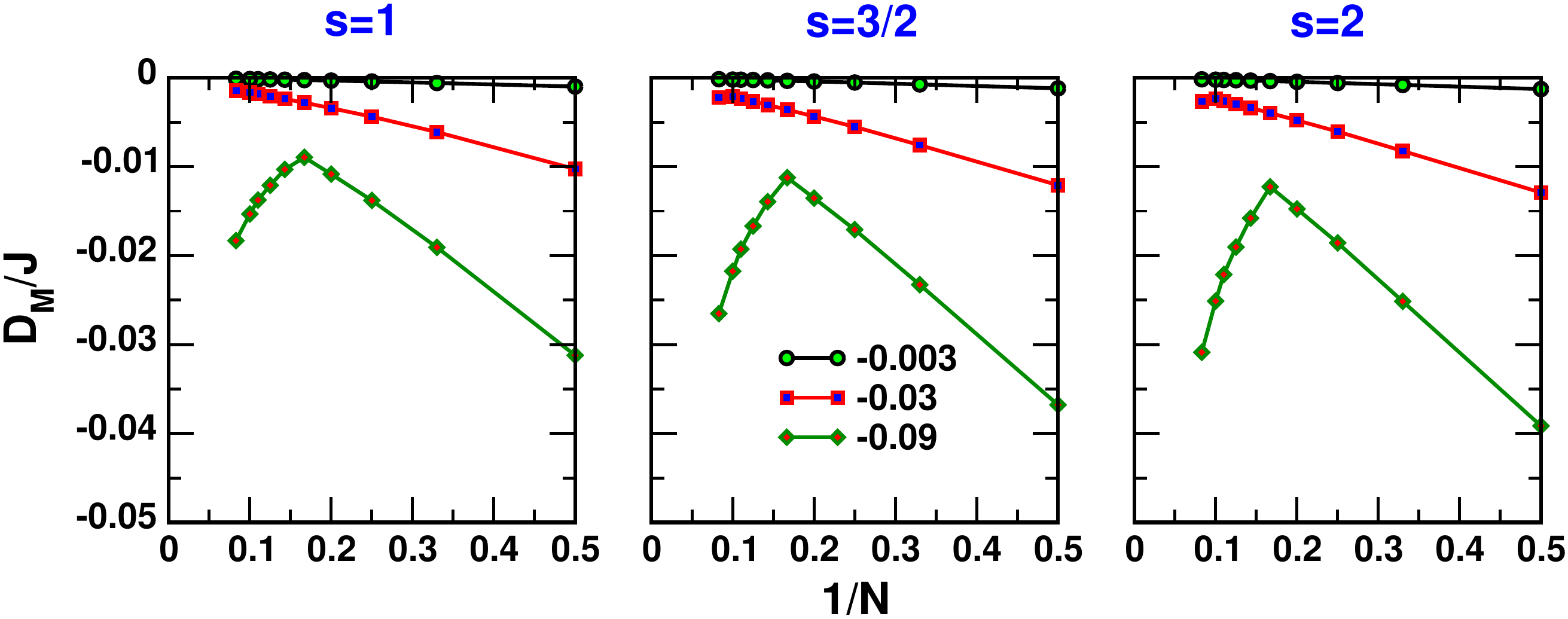}
    \caption{\label{fig:DMvs1byN_WA}Effect of system size $1/N$ on the molecular anisotropy $D_M$ of ring for $d=$-0.003, -0.03 and 0.09J.}
\end{figure*}

\begin{figure*}
    \includegraphics[width=\textwidth]{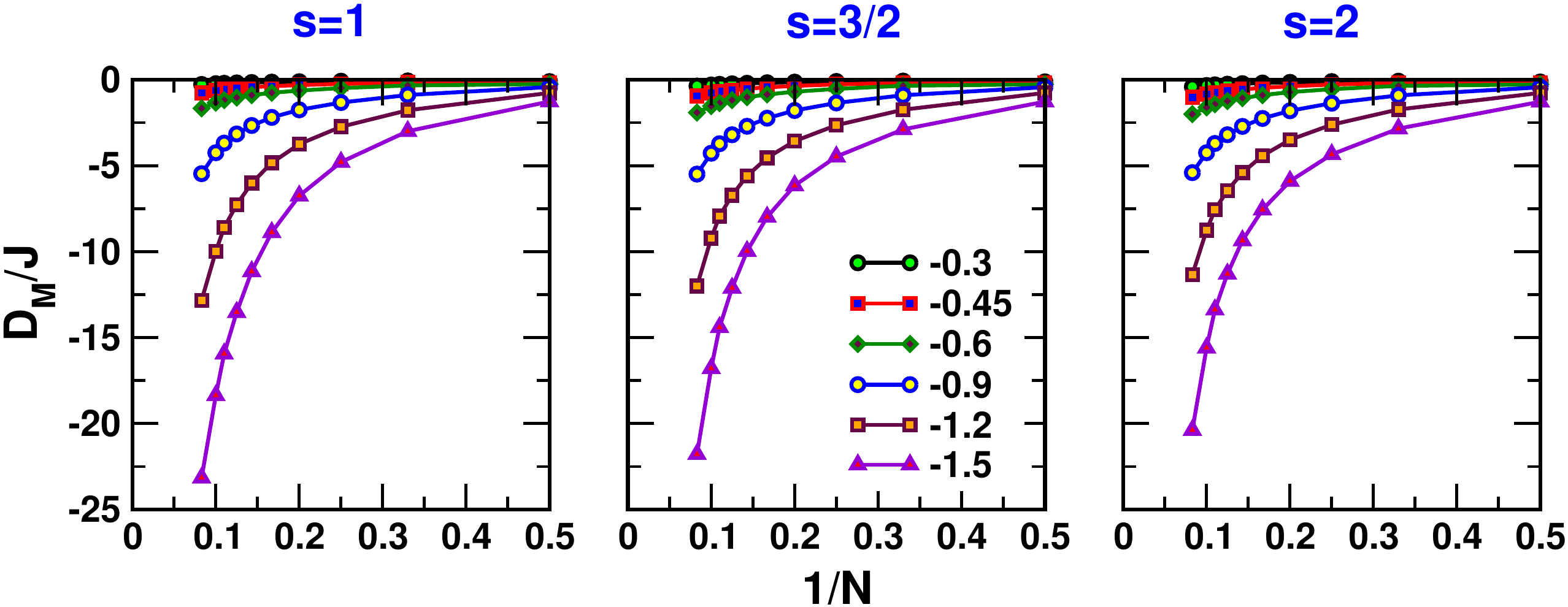}
    \caption{\label{fig:DMvs1byN_SA}Effect of system size $1/N$ on the molecular anisotropy $D_M$ of ring for $d=$-0.3, -0.45, -0.6 and -0.9J.}
\end{figure*}
